\documentclass[a4paper]{article}

\usepackage[margin=1in]{geometry}

\usepackage[T1]{fontenc}
\newcommand{\changefont}[3]{
\fontfamily{#1} \fontseries{#2} \fontshape{#3} \selectfont}

\changefont{ptm}{m}{n}

\usepackage{setspace} 
\usepackage{graphicx}    % needed for including graphics e.g. EPS, PS

\usepackage{amsfonts,amssymb,amsmath}
\usepackage{graphics}
\usepackage{mathrsfs}
\usepackage{color}

\newtheorem{theorem}{Theorem}[section]

\newtheorem{definition}{Definition}[section]

\long\def\symbolfootnote[#1]#2{\begingroup%
\def\thefootnote{\fnsymbol{footnote}}\footnote[#1]{#2}\endgroup} 

\begin{document}

\begin{center}
\Large \textbf{Unpredictable sequences and Poincar\'e chaos}
\end{center}

\begin{center}
\normalsize \textbf{Marat Akhmet$^{1,}$\symbolfootnote[1]{Corresponding Author. E-mail: marat@metu.edu.tr, Tel: +90 312 210 5355}, Mehmet Onur Fen$^2$} \\
\vspace{0.2cm}
\textit{\textbf{$^1$Department of Mathematics, Middle East Technical University, 06800 Ankara, Turkey}}

\vspace{0.1cm}
\textit{\textbf{$^2$Basic Sciences Unit, TED University, 06420 Ankara, Turkey}}
\vspace{0.1cm}
\end{center}

\vspace{0.3cm}

\begin{center}
\textbf{Abstract}   
\end{center}

\vspace{-0.2cm}

\noindent\ignorespaces

To make research of chaos more friendly with discrete equations,  we introduce the concept of an unpredictable sequence as a specific unpredictable function on the set of integers. It is convenient to be verified as a solution of a discrete equation. This is rigorously proved in this paper for quasilinear systems, and we demonstrate the result numerically for linear systems in the critical case with respect to the stability of the origin. The completed research contributes to the theory of chaos as well as to the theory of discrete equations, considering unpredictable solutions. 

\vspace{0.2cm}
 
\noindent\ignorespaces \textbf{Keywords:} Unpredictable sequences; Unpredictable solutions; Quasilinear systems; Poincar\'e chaos.

\vspace{0.6cm}

%%%%%%%%%%%%%%%%%%%%%%%%%%%%%%%%%%%%%%%%%%%%%%%%%%%%%%%%%%%%%%%%%%%%%%%%%%%%%%%%%%%%%%%%%%%%%% 

\section{Introduction}

In the  instrumental  sense, discreteness has been the main object in  chaos  investigation. To check this, it is sufficient to recall definitions of chaos \cite{ly}-\cite{Wiggins88}. They are based on sequences and maps. One can say that stroboscopic observation of a motion was the single way to indicate the irregularity in a continuous dynamics. Embedding the research to the theory of differential equations requests definitions of chaos for continuous dynamics, which are not related to discreteness \cite{Akh14}-\cite{Akh17}. The research as well as the origin of the chaos \cite{poin} gave us strong arguments for the development of motions in classical dynamical systems theory \cite{bir} by proceeding behind Poisson stable points to unpredictable points \cite{Akh21}. Then, the dynamics has been specified such that a function that is bounded on the real axis is an unpredictable point \cite{Akh24}-\cite{Akh22}. In the papers \cite{Akh24}-\cite{Akh22}, we have demonstrated that unpredictable functions are easy to be analyzed as solutions of differential equations. This paradigm is not completed, if one does not consider discrete equations. This is the reason why in the present paper we deliver discrete analogues for unpredictable functions calling them unpredictable sequences, and prove for the first time in the literature assertions on the existence and uniqueness of unpredictable solutions of recurrent equations. The role of the novelty cannot be underestimated for applications as well as for theoretical analysis, exceptionally for the modern development of computer technologies, software, and robotics \cite{ditto,sta}.  

\section{Preliminaries} \label{disc_sec_prelim}

Throughout the paper, we will make use of the usual Euclidean norm for vectors and the norm induced by the Euclidean norm for matrices.

Let $(X, d)$ be a metric space, and $\mathbb T$ refer to either the set of real numbers or the set of integers. Suppose that $\pi: \mathbb T \times X \to X$ is a flow on $X$, i.e., 
$\pi(0,\sigma)=\sigma$ for all $\sigma \in X,$
$\pi(t,\sigma)$ is continuous in the pair of variables $t$ and $\sigma,$ and
$\pi(t_1, \pi(t_2,\sigma))=\pi(t_1+t_2,\sigma)$ for all $t_1,$ $t_2 \in \mathbb T$ and $\sigma \in X$  \cite{sell}. 
We modified the Poisson stable points to unpredictable points in paper \cite{Akh21} as follows.  

\begin{definition}  \label{SSP_point}
A point $\sigma \in X$ and the trajectory through it are \textit{unpredictable} if there exist a positive number $\epsilon_0$ (the unpredictability constant) and sequences $\left\{t_k\right\}$  and  $\left\{\tau_k\right\},$ both of which diverge to infinity, such that
$\displaystyle \lim_{k \to \infty} \pi(t_k,\sigma)=\sigma$ and $d[\pi(t_k+\tau_k,\sigma), \pi(\tau_k,\sigma)] \ge \epsilon_0$ for each $k \in \mathbb N.$
\end{definition}

To develop the row of periodic, quasiperiodic, and almost periodic oscillations to a new one, we specified in \cite{Akh24}-\cite{Akh22} unpredictability for functions as points of a dynamics.

\begin{definition} \label{definition_unpredictable}
A uniformly continuous and bounded function $\varphi: \mathbb R  \to  \mathbb R^p$ is unpredictable if there exist positive numbers $\epsilon_0,$ $\delta,$ and sequences $\left\{t_k\right\},$ $\left\{\tau_k\right\}$, both of which diverge to infinity, such that $\left\|\varphi(t+t_k)-\varphi(t)\right\| \to 0$ as $k \to \infty$ uniformly on compact subsets of $\mathbb R,$ and $\left\|\varphi(t+t_k)-\varphi(t)\right\| \ge \epsilon_0$ for each $t\in [\tau_k-\delta, \tau_k+\delta]$ and $n\in\mathbb N.$
\end{definition}

The last definition can be considered as a more restrictive version of the next two, which will be useful in the future for applications of functional analysis methods in the theory of differential equations.

\begin{definition} \label{definition_unpredictable1}
A continuous and bounded function $\varphi: \mathbb R \to \mathbb R^p$ is unpredictable if there exist positive numbers $\epsilon_0,$ $\delta,$ and sequences $\left\{t_k\right\},$ $\left\{\tau_k\right\}$, both of which diverge to infinity, such that $\left\|\varphi(t+t_k)-\varphi(t)\right\| \to 0$ as $k \to \infty$ uniformly on compact subsets of $\mathbb R$ and $\left\|\varphi(t_k+\tau_k)-\varphi(\tau_k)\right\| \ge \epsilon_0$ for each $k\in\mathbb N.$
\end{definition}

\begin{definition} \label{definition_unpredictable2}
A continuous and bounded function $\varphi: \mathbb R \to \mathbb R^p$ is unpredictable if there exist positive numbers $\epsilon_0,$ $\delta,$ and sequences $\left\{t_k\right\},$ $\left\{\tau_k\right\}$, both of which diverge to infinity, such that $\left\|\varphi(t_k)-\varphi(0)\right\| \to 0$ as $k \to \infty$ and $\left\|\varphi(t_k+\tau_k)-\varphi(\tau_k)\right\| \ge \epsilon_0$ for each $k\in\mathbb N.$
\end{definition}

The following definition of an unpredictable sequence was first mentioned in paper \cite{Akh22} as an analogue for Definition \ref{definition_unpredictable}.

\begin{definition} \label{unp_seq_defn}
A bounded sequence $\left\{\varphi_n\right\},$ $n\in\mathbb Z,$ in $\mathbb R^p$ is called unpredictable if there exist a positive number $\epsilon_{0}$ and  sequences $\left\{\zeta_k\right\},$ $\left\{\eta_k\right\}$, $k\in\mathbb N$, of positive integers, both of which diverge to infinity, such that $\left\|\varphi_{n+\zeta_k} - \varphi_n \right\|\to 0$ uniformly as $k \to \infty$ for each $n$ in bounded intervals of integers and $ \left\|\varphi_{\zeta_k + \eta_k} - \varphi_{\eta_k}\right\| \geq \epsilon_{0}$ for each $k\in\mathbb N.$
\end{definition}

Definition \ref{unp_seq_defn} is of main use in the present paper. It is requested by the method of the proof. Nevertheless, in future analyses, there may be needs for another definition, which can be considered as a direct specification of Definition \ref{SSP_point} as well as an  analogue of Definition \ref{definition_unpredictable2}. 

\begin{definition} \label{unp_seq_defn2}
A bounded sequence $\left\{\varphi_n\right\},$ $n\in\mathbb Z,$ in $\mathbb R^p$ is called unpredictable if there exist a positive number $\epsilon_{0}$ and  sequences $\left\{\zeta_k\right\},$ $\left\{\eta_k\right\}$, $k\in\mathbb N$, of positive integers, both of which diverge to infinity, such that $\left\|\varphi_{\zeta_k} - \varphi_0 \right\|\to 0,$  as $k \to \infty,$ and $ \left\|\varphi_{\zeta_k + \eta_k} - \varphi_{\eta_k}\right\| \geq \epsilon_{0}$ for each $k\in\mathbb N.$
\end{definition}

The topologies in Definitions \ref{definition_unpredictable} and \ref{unp_seq_defn} are metrizable \cite{sell}. Consequently, the existence of an unpredictable sequence in the sense of Definition \ref{unp_seq_defn} indicates the presence of Poincar\'e chaos \cite{Akh21}. In what  follows, an unpredictable sequence and an unpredictable solution are understood as mentioned in Definition \ref{unp_seq_defn}.  

In this paper, we will consider the following discrete equation,
\begin{eqnarray} \label{main_discrete_eqn}
x_{n+1} = A x_{n} + f(x_n) + \psi_n,
\end{eqnarray}
where $n\in\mathbb Z,$ $A\in\mathbb R^{p \times p}$ is a nonsingular matrix, $f:\mathbb R^p \to \mathbb R^p$ is a continuous function, and $\left\{\psi_n\right\},$ $n\in\mathbb Z,$ is an unpredictable sequence.

The following assumptions on equation (\ref{main_discrete_eqn}) are required.
\begin{enumerate}
\item[\bf (C1)] There exists a positive number $M_f$ such that $\displaystyle \sup_{x\in\mathbb R^p} \left\|f(x)\right\| \leq M_f;$ 
\item[\bf (C2)] There exists a positive number $L_f$ such that $\left\|f(x)-f(y)\right\|\leq L_f \left\|x-y\right\|$ for all $x$, $y\in\mathbb R^p;$
\item[\bf (C3)] $\left\|A\right\| + L_f < 1$.
\end{enumerate}

According to the results of \cite{Laksh02}, if conditions $(C1)-(C3)$ hold, then equation (\ref{main_discrete_eqn}) possesses a unique bounded solution $\left\{\phi_n\right\}$, $n\in\mathbb Z,$ which satisfies the relation
\begin{eqnarray} \label{discrete_bdd_soln}
\phi_n = \sum_{j=-\infty}^n A^{n-j} \left[f(\phi_{j-1}) + \psi_{j-1}\right].
\end{eqnarray}
One can show under the same conditions that the bounded solution attracts all other solutions of (\ref{main_discrete_eqn}). More precisely, the inequality 
$$
\left\|x_n-\phi_n\right\| \leq \left(\left\|A\right\|+L_f\right)^{(n-n_0)} \left\|x^{0}-\phi_{n_0}\right\|
$$
is valid for all $n\geq n_0$, where $\left\{x_n\right\}$, $n\in\mathbb Z$, is a solution of (\ref{main_discrete_eqn}) with $x_{n_0}=x^0$ for some integer $n_0$ and $x^0\in \mathbb R^p$.

\section{Unpredictable sequences} \label{disc_sec_uprdsec}

The following theorem is concerned with the existence of an unpredictable solution of the discrete equation (\ref{main_discrete_eqn}).

\begin{theorem} \label{main_thm_discrete}
The bounded solution $\left\{\phi_n\right\},$ $n\in\mathbb Z,$ of equation (\ref{main_discrete_eqn}) is unpredictable under the conditions $(C1)-(C3)$.
\end{theorem}

\noindent \textbf{Proof.} Fix an arbitrary positive number $\epsilon,$ and suppose that $\gamma$ is a positive number satisfying 
$$
\gamma \leq \left[\frac{1}{1-\left\|A\right\|-L_f} + \frac{2(M_f+M_{\psi})}{1-\left\|A\right\|}\right]^{-1}.
$$
Let $n_1$ and $n_2$ be integers such that $n_2>n_1$, and take a natural number $E$ with 
\begin{eqnarray} \label{discr_numberE}
E\geq \displaystyle \frac{\ln (\gamma \epsilon)}{\ln(\left\|A\right\|+L_f)} -1.
\end{eqnarray}
Since $\left\{\psi_n\right\},$ $n\in\mathbb Z,$ is an unpredictable sequence, there exist a positive number $\epsilon_0$ and sequences $\left\{\zeta_k\right\},$ $\left\{\eta_k\right\}$, $k\in\mathbb N$, of positive integers both of which diverge to infinity such that $\left\|\psi_{n+\zeta_k} - \psi_n \right\|\to 0$ uniformly as $k \to \infty$ for each $n$ with $n_1-E-1 \leq n \leq n_{2}-1$ and $ \left\|\psi_{\zeta_k + \eta_k} - \psi_{\eta_k}\right\| \geq \epsilon_{0}$ for each $k\in\mathbb N.$

First of all, we will show that $\left\|\phi_{n+\zeta_k} - \phi_n \right\|\to 0$ uniformly as $k \to \infty$ for each $n$ with $n_{1} \leq n \leq n_{2}$.
There exists a natural number $k_0,$ independent of $n,$ such that for each $k \geq k_0$ the inequality $\left\|\psi_{n+\zeta_k} - \psi_n \right\| < \gamma \epsilon$ is valid whenever $n_1-E-1 \leq n \leq n_2-1$.

Fix an arbitrary integer $k \geq k_0$. One can obtain using the relation (\ref{discrete_bdd_soln}) that
\begin{eqnarray*}
\phi_{n+\zeta_k} - \phi_n = \displaystyle \sum_{j=-\infty}^{n} A^{n-j} \left[ f(\phi_{j+\zeta_k-1}) - f(\phi_{j-1}) + \psi_{j+\zeta_k-1} - \psi_{j-1}  \right].
\end{eqnarray*}
Therefore, for $n_1-E \leq n \leq n_2,$ we have
\begin{eqnarray}\label{disc_ineq1_proof}
\begin{array}{l}
\displaystyle \left\|\phi_{n+\zeta_k}-\phi_n\right\|  < \displaystyle \frac{2(M_{f}+M_{\psi})}{1-\left\|A\right\|} \left\|A\right\|^{n-n_1+E+1} + \frac{\gamma \epsilon}{1-\left\|A\right\|} \left(1-\left\|A\right\|^{n-n_1+E+1}\right) \\
 \ \ \ \ \ \  \ \ \ \ \ \   \ \ \ \ \ \  \   + L_{f} \displaystyle \sum_{j=n_1-E}^n \left\|A\right\|^{n-j} \left\|\phi_{j+\zeta_k-1} - \phi_{j-1}\right\|.
\end{array}
\end{eqnarray} 
Let us denote 
$$
z_n=\left\|A\right\|^{-n} \left\|\phi_{n+\zeta_k}-\phi_n\right\|
$$
and
$$
\vartheta_n = \displaystyle \frac{2(M_{f}+M_{\psi})}{1-\left\|A\right\|} \left\|A\right\|^{-n_1+E+1} + \frac{\gamma \epsilon}{1-\left\|A\right\|} \left(\left\|A\right\|^{-n} - \left\|A\right\|^{-n_1+E+1}\right).
$$
The inequality (\ref{disc_ineq1_proof}) yields
\begin{eqnarray*}
z_n < \vartheta_n + \frac{L_{f}}{\left\|A\right\|} \displaystyle \sum_{j=n_1-E}^n z_{j-1}.
\end{eqnarray*}
It can be verified by applying the discrete analogue of Gronwall inequality that
$$
z_n \leq \vartheta_n + \displaystyle  \frac{L_{f}}{\left\|A\right\|} \sum_{j=n_1-E}^{n} \vartheta_{j-1} \left(1+\frac{L_f}{\left\|A\right\|}\right)^{n-j}.
$$
Thus, for $n_1-E\leq n \leq n_2$, we have
\begin{eqnarray*}
& z_n & \leq \displaystyle \frac{2(M_{f}+M_{\psi})}{1-\left\|A\right\|}  \left\|A\right\|^{-n} \left(\left\|A\right\|+L_{f}\right)^{n-n_1+E+1}\\
&& + \frac{\gamma \epsilon}{1-\left\|A\right\|-L_f}  \left\|A\right\|^{-n}  \left[1-\left(\left\|A\right\|+L_{f}\right)^{n-n_1+E+1}\right].
\end{eqnarray*}
The last inequality implies that    
\begin{eqnarray*}
\left\|\phi_{n+\zeta_k} - \phi_{n}\right\| <  \frac{2(M_{f}+M_{\psi})}{1-\left\|A\right\|} \left(\left\|A\right\|+L_{f}\right)^{n-n_1+E+1} + \displaystyle \frac{\gamma \epsilon}{1-\left\|A\right\|-L_{f}}.
\end{eqnarray*}
One can confirm using (\ref{discr_numberE}) that $\left(\left\|A\right\|+L_{f}\right)^{n-n_1+E+1} \leq \gamma\epsilon$ for $n_1 \leq n \leq n_2$. Therefore, for each $k \geq k_0$, the inequality
$$
\left\|\phi_{n+\zeta_k} - \phi_{n}\right\| < \left[ \frac{2(M_f+M_{\psi})}{1-\left\|A\right\|} + \frac{1}{1-\left\|A\right\|-L_f} \right] \gamma \epsilon \leq \epsilon
$$
is valid for $n_1 \leq n \leq n_2$. Hence, $\left\|\phi_{n+\zeta_k} - \phi_n \right\|\to 0$ uniformly as $k \to \infty$ for each $n$ with $n_{1} \leq n \leq n_{2}$.

Next, we will show the existence of a positive number $\overline{\epsilon}_0$ and a sequence $\left\{\widetilde{\eta}_k\right\}$ with $\widetilde{\eta}_k \to \infty$ as $k \to \infty$ such that $\left\|\phi_{\zeta_{k}+\widetilde{\eta}_{k}} - \phi_{\widetilde{\eta}_k}\right\| \geq \overline{\epsilon}_0$ for each $k \in \mathbb N$.

Using the relations
$$
\phi_{\zeta_k+\eta_k+1} = A \phi_{\zeta_k+\eta_k} + f(\phi_{\zeta_k+\eta_k}) + \psi_{\zeta_k+\eta_k}
$$
and
$$
\phi_{\eta_k+1}=A\phi_{\eta_k} + f(\phi_{\eta_k}) + \psi_{\eta_k}
$$
we obtain for $k\in\mathbb N$ that
$$
\left\|\phi_{\zeta_k+\eta_k+1} - \phi_{\eta_k+1}\right\| \geq \epsilon_0 - \left(\left\|A\right\|+L_{f}\right) \left\|\phi_{\zeta_k+\eta_k} - \phi_{\eta_k}\right\|.
$$
Therefore,
\begin{eqnarray} \label{discrete_ineq_2}
\max\left\{ \left\|\phi_{\zeta_k+\eta_k+1} - \phi_{\eta_k+1}\right\|,  \left\|\phi_{\zeta_k+\eta_k} - \phi_{\eta_k}\right\| \right\} \geq \overline{\epsilon}_0,
\end{eqnarray}
where $\overline{\epsilon}_0=\displaystyle \frac{\epsilon_0}{1+\left\|A\right\|+L_{f}}$.

For each $k\in\mathbb N$, let us take $\widetilde{\eta}_k=\eta_{k}+1$ if $\left\|\phi_{\zeta_k+\eta_k+1} - \phi_{\eta_k+1}\right\| \geq \left\|\phi_{\zeta_k+\eta_k} - \phi_{\eta_k}\right\|$, and we set $\widetilde{\eta}_k=\eta_k$ otherwise. Clearly, $\widetilde{\eta}_k \to \infty$ as $k\to\infty$. According to inequality (\ref{discrete_ineq_2}), we have $\left\|\phi_{\zeta_k+\widetilde{\eta}_k} - \phi_{\widetilde{\eta}_k} \right\| \geq \overline{\epsilon}_0$ for each $k\in\mathbb N.$ 
Consequently, the bounded solution $\left\{\phi_n\right\},$ $n\in\mathbb Z,$ of (\ref{main_discrete_eqn}) is unpredictable. $\square$

A possible way to obtain a different unpredictable sequence from a given one is mentioned in the following theorem.

\begin{theorem} \label{discrete_thm2}
Suppose that $\left\{\varphi_n\right\}$, $n\in\mathbb Z$, is an unpredictable sequence such that $\varphi_n \in \Lambda$ for each $n$, where $\Lambda$ is a bounded subset of $\mathbb R^p$. If $h:\Lambda \to \mathbb R^q$ is a function such that there exist positive numbers $L_1$ and $L_2$ with $L_1\left\|s_1-s_2\right\|\leq \left\|h(s_1)-h(s_2)\right\|\leq L_2\left\|s_1-s_2\right\|$ for all $s_1,$ $s_2\in \Lambda$, then the sequence $\left\{\overline{\varphi}_n\right\}$ defined through the equation $\overline{\varphi}_n=h(\varphi_n)$, $n \in \mathbb Z$, is also unpredictable.
\end{theorem}

\noindent \textbf{Proof.} Since $\left\{\varphi_n\right\}$, $n\in\mathbb Z$, is an unpredictable sequence, there exist a positive number $\epsilon_{0}$ and sequences $\left\{\zeta_k\right\},$ $\left\{\eta_k\right\}$, $k\in\mathbb N$, of positive integers both of which diverge to infinity such that $\left\|\varphi_{n+\zeta_k} - \varphi_n \right\|\to 0$ uniformly as $k \to \infty$ for each $n$ in bounded intervals of integers and $ \left\|\varphi_{\zeta_k + \eta_k} - \varphi_{\eta_k}\right\| \geq \epsilon_{0}$ for each $k\in\mathbb N.$

Fix an arbitrary positive number $\epsilon,$ and let $n_1$ and $n_2$ be any two integers such that $n_2>n_1$. One can find a natural number $k_0$, which does not depend on $n$, such that for each $k\geq k_0$ we have $\left\|\varphi_{n+\zeta_k}-\varphi_n\right\|<\epsilon/L_2$ whenever $n_1\leq n\leq n_2$. Therefore, the inequality
$$
\left\|\overline{\varphi}_{n+\zeta_k} - \overline{\varphi}_n\right\| \leq L_2 \left\|\varphi_{n+\zeta_k}-\varphi_n\right\| <\epsilon
$$
is satisfied for each $k\geq k_0$ and each $n$ with $n_1 \leq n \leq n_2$. This shows that $\left\|\overline{\varphi}_{n+\zeta_k} - \overline{\varphi}_n\right\| \to 0$ uniformly as $k\to \infty$ on bounded intervals of integers. On the other hand, for each $k\in\mathbb N$, we have that
$$
\left\|\overline{\varphi}_{\zeta_k+\eta_k} - \overline{\varphi}_{\eta_k}\right\| \geq L_1 \left\|\varphi_{\zeta_k+\eta_k} - \varphi_{\eta_k} \right\| \geq L_1 \epsilon_0.
$$
Consequently, $\left\{\overline{\varphi}_n\right\}$, $n\in\mathbb Z$, is an unpredictable sequence. $\square$

In the next section, an example which supports the result of Theorem \ref{disc_sec_uprdsec} is provided.

\section{An example} \label{discrete_example}

Consider the logistic map
\begin{eqnarray} \label{disc_log_map}
u_{n+1} = \lambda u_n (1-u_n),
\end{eqnarray}
where $n\in\mathbb Z$ and $\lambda$ is a parameter. Based on the result of the papers \cite{Akh23} and \cite{Shi07}, it was demonstrated in paper \cite{Akh22} that for $\lambda \in [3+(2/3)^{1/2},4]$ the map (\ref{disc_log_map}) possesses an unpredictable solution. For such values of the parameter, the unit interval $[0,1]$ is invariant under the iterations of the map \cite{Hale91}.

Next, we take into account the discrete system
\begin{eqnarray} \label{example_discrete_system}
\begin{array}{l}
x_{n+1}= \displaystyle \frac{x_n}{2} - \frac{y_n}{7} + 3\psi_n^3 \\
y_{n+1}= - \displaystyle \frac{x_n}{8} + \frac{y_n}{3} + \frac{x_n^{2/3}}{12} + 4\psi_n,
\end{array}
\end{eqnarray}
where $\left\{\psi_n\right\}$ is an unpredictable solution of (\ref{disc_log_map}) with $\lambda=3.91$. Theorem \ref{discrete_thm2} implies that the sequence $\left\{\overline{\psi}_n\right\}$, $n\in\mathbb Z$,  defined as $\overline{\psi}_n=(3\psi_n^3, \ 4\psi_n)\in\mathbb R^2$ is also unpredictable.

In order to demonstrate the chaotic behavior of (\ref{example_discrete_system}), we consider the system
\begin{eqnarray} \label{example_discrete_system2}
\begin{array}{l}
x_{n+1}= \displaystyle \frac{x_n}{2} - \frac{y_n}{7} + 3u_n^3 \\
y_{n+1}= - \displaystyle \frac{x_n}{8} + \frac{y_n}{3} + \frac{x_n^{2/3}}{12} + 4u_n,
\end{array}
\end{eqnarray}
where $\left\{u_n\right\}$ is a solution of (\ref{disc_log_map}), again with $\lambda=3.91$. One can numerically verify that for each $\left\{u_n\right\}$, the bounded solutions of (\ref{example_discrete_system2}) take place inside the compact region
$$
\mathcal{D}=\{(x,y)\in\mathbb R^2: \ 0.1 \leq x \leq 2.7, \ 1.7 \leq y \leq 5.1\}.
$$
Therefore, the conditions $(C1)-(C3)$ are satisfied for system (\ref{example_discrete_system}), and there exists a unique unpredictable solution of (\ref{example_discrete_system}) in accordance with Theorem \ref{main_thm_discrete}.

\begin{figure}[ht] 
\centering
\includegraphics[width=13.5cm]{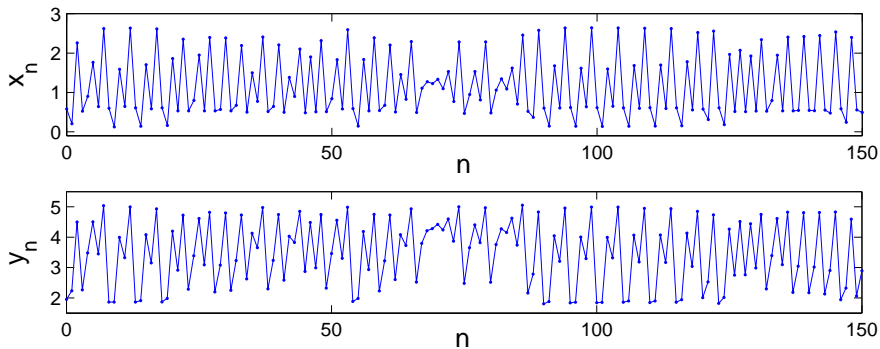}
\caption{The solution of (\ref{example_discrete_system2}) with the initial data $u_0=0.4,$ $x_0=0.58,$ and $y_0=1.95$. The figure supports the result of Theorem \ref{main_thm_discrete} such that (\ref{example_discrete_system}) possesses an unpredictable solution.}
\label{fig1}
\end{figure}

\begin{figure}[ht] 
\centering
\includegraphics[width=7.0cm]{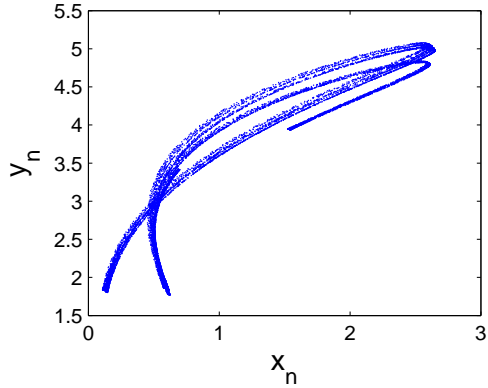}
\caption{The trajectory of the discrete system (\ref{example_discrete_system2}) corresponding to the initial data $u_0=0.4,$ $x_0=0.58,$ and $y_0=1.95$. The figure reveals the chaotic behavior of system (\ref{example_discrete_system}).}
\label{fig2}
\end{figure}

Figure \ref{fig1} shows the first and second coordinates of the  solution of system (\ref{example_discrete_system2}) with the initial data $u_0=0.4,$ $x_0=0.58,$ and $y_0=1.95$. The utilized value of the parameter $\lambda=3.91$ and the initial point $u_0=0.4$ were considered for shadowing in paper \cite{Hammel87}. Moreover, we represent in Figure \ref{fig2} the two dimensional trajectory of the same solution. Both Figures \ref{fig1} and \ref{fig2} support the result of Theorem \ref{main_thm_discrete} such that an unpredictable sequence takes place in the dynamics of the discrete system (\ref{example_discrete_system}) and the behaviour of the system is chaotic.

\section{Poincar\'e chaos near periodic orbits} \label{discrete_periodic}

In this section, we will demonstrate the appearance of irregular behavior near periodic orbits of discrete systems. For that purpose, let us consider the system
\begin{eqnarray} \label{discrete_per}
\begin{array}{l}
x_{n+1}= \cos(\omega) x_n + \sin(\omega) y_n \\
y_{n+1}= - \sin(\omega) x_n + \cos(\omega) y_n.
\end{array}
\end{eqnarray}
It is shown in the book \cite{Hale91} that the system (\ref{discrete_per}) admits a stable periodic orbit whenever the value $\omega/2\pi$ is rational. 
Taking $\lambda=3.86$ in the logistic map (\ref{disc_log_map}), and perturbing system (\ref{discrete_per}) with solutions (\ref{disc_log_map}) we set up the system
\begin{eqnarray} \label{discrete_per2}
\begin{array}{l}
x_{n+1}= \cos(\omega) x_n + \sin(\omega) y_n + 0.001 u_n \\
y_{n+1}= - \sin(\omega) x_n + \cos(\omega) y_n + 0.001 u_n
\end{array}
\end{eqnarray}
where $\left\{u_n\right\}$ is a solution of (\ref{disc_log_map}).

\begin{figure}[ht] 
\centering
\includegraphics[width=7.4cm]{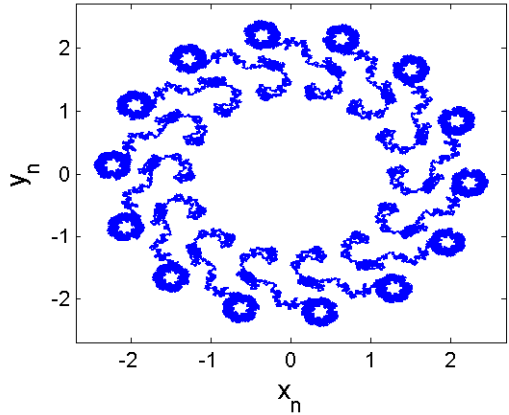}
\caption{The orbit of system (\ref{discrete_per2}) with $u_0=0$, $x_0=1$, and $y_0=1$. The figure manifests that the orbit behaves chaotically near the $14$-periodic orbit of (\ref{discrete_per}).}
\label{fig3}
\end{figure}

Let us use the value of $\omega=\pi/7$ so that the non-perturbed system (\ref{discrete_per}) possesses a one parameter family of stable $14$-periodic orbits. We depict in Figure \ref{fig3} the trajectory of (\ref{discrete_per2}) corresponding to the initial data $u_0=0.4$, $x_0=1$, and $y_0=1$. The total number of iterations used in the simulation is $65 \times 10^6$. The choice for the parameter value $\lambda=3.86$ and the initial value $u_0=0.4$ is analyzed for shadowing in paper \cite{Hammel87}. It is seen in Figure \ref{fig3} that the applied perturbation makes the system (\ref{discrete_per2}) behave chaotically near the $14$-periodic orbit of (\ref{discrete_per}). It is worth noting that Figure \ref{fig3} represents a single orbit. The fractal structure of the orbit is also observable in the simulation. Figure \ref{fig3} manifests the appearance of Poincar\'e chaos near the periodic orbit of (\ref{discrete_per}).

\section{Conclusion}
The starting point for the present research is the unpredictable point \cite{Akh21}, a new object for the dynamical systems theory founded by Poincar\'e and Birkhoff \cite{poin,bir}. In the paper \cite{Akh21}, we developed the Poisson stability of a motion to unpredictability such that a new type of chaos, the Poincar\'e chaos, has been obtained.  It has become clear that the concept can be easily extended to the object of analysis in the theory of differential equations, considering unpredictable functions as points moving by shifts of the time argument \cite{Akh24}-\cite{Akh22}. Therefore, in our opinion, a new field to analyze in the theory of differential equations has been discovered. Since many results of differential equations have their counterparts in discrete equations, it is easy to suppose that theorems on the existence of unpredictable solutions can be proved for discrete equations. The present paper is the first one to realize the paradigm. The existence and uniqueness theorem for quasilinear difference equations has been proved, when the  perturbation is an unpredictable sequence. This is visualized as Poincar\'e chaos in simulations.

\end{document}